\documentclass[aps,prb,twocolumn,10pt,showpacs,amsmath,amssymb]{revtex4-1}
\usepackage{graphicx}

\begin{document}

\title{Unconventional magnetism in multivalent charge-ordered YbPtGe$_2$ \\
probed by $^{195}$Pt- and $^{171}$Yb-NMR}

\author{R. Sarkar}
\email{rajibsarkarsinp@gmail.com}
\altaffiliation[present address: ]{Institute for Solid State Physics,
TU Dresden, 01069 Dresden, Germany}
\author{R. Gumeniuk}
\author{A. {Leithe-Jasper}}
\author{W. Schnelle}
\author{Y. Grin}
\author{C. Geibel}
\author{M. Baenitz}
\affiliation{Max-Planck Institute for Chemical Physics of Solids,
01187 Dresden, Germany}

\date{\today}

\begin{abstract}
Detailed $^{195}$Pt- and $^{171}$Yb nuclear magnetic resonance (NMR) studies
on the heterogeneous mixed valence system YbPtGe$_2$ are reported. The
temperature dependence of the $^{195}$Pt-NMR shift $^{195}K(T)$ indicates the
opening of an unusual magnetic gap below 200\,K. $^{195}K(T)$ was analyzed by
a thermal activation model which yields an isotropic gap $\Delta/k_B \approx
200$\,K. In contrast, the spin-lattice relaxation rate $^{195}$($1/T_1$) does
not provide evidence for the gap. Therefore, an intermediate-valence picture
is proposed while a Kondo-insulator scenario can be excluded. Moreover,
$^{195}$($1/T_1$) follows a simple metallic behavior, similar to the reference
compound YPtGe$_2$. A well resolved NMR line with small shift is assigned to
divalent $^{171}$Yb. This finding supports the proposed model with two
sub-sets of Yb species (di- and trivalent) located on the Yb2 and Yb1 site of
the YbPtGe$_2$ lattice.
\end{abstract}


\maketitle

Intermetallic compounds formed by $d$-transition metals with mostly itinerant
electrons and rare-earth metals with more localized electrons ($4f$-electrons)
have been the subject of intense research activities in the solid state
science. The importance of these systems is the presence of an effective
hybridization between localized and itinerant electrons giving rise to exotic
properties such as, mixed- or intermediate-valence states, heavy-fermion and
non-Fermi liquid behaviors, unconventional superconductivity, Kondo
insulators, Kondo semiconductors, quantum
criticality.\cite{Adroja-2008,Riseborough-2000,Stewart-2001,Georges-1996}

Among the properties listed above the heavy-fermion semiconductors,
alternately called Kondo insulators, attracted much attention in the recent
years. These materials exhibit a small energy gap at the Fermi level, which is
believed to appear due to the hybridization between localized $d$- or
$f$-electrons with conduction electrons. Kondo insulators associated with such
a hybridization gap are characterized by an insulating or semiconducting
electrical resistivity, a non-magnetic ground state, and local-moment
magnetism at temperatures far above the gap opening. The best known examples
of such materials are YbB$_{12}$,\cite{Kasaya-1983,Iga-1988,Kasaya-1985}
Ce$_3$Bi$_4$Pt$_3$,\cite{Hundley-1990,Jaime-2000}
FeSi,\cite{Riseborough-2000,Jaccarino-1967}
U$_2$Ru$_2$Sn,\cite{Rajarajan-2007} CeRu$_4$Sn$_6$,\cite{Bruening-ceru4sn6}
and CeFe$_2$Al$_{10}$.\cite{Lue-PRB-CeFe2Al10-2010} There are also Kondo
insulators which exhibit metal-like electrical resistivity at low
temperatures, caused by residual ``in-gap states'' and an anisotropic Fermi
surface. Typical examples are CeNiSn,\cite{Takabatake-1990}
CeRu$_4$Sn$_6$,\cite{Bruening-ceru4sn6} and SmB$_6$.\cite{Canfield-1995-SmB6}
Just recently these ``in-gap states'' got an revived attention because the
question arises whether such states are actually metallic surfaces of a
three-dimensional topological insulator.\cite{Dzero-2010a,Dzero-2012a}

\begin{figure}
\includegraphics[width=3.0in]{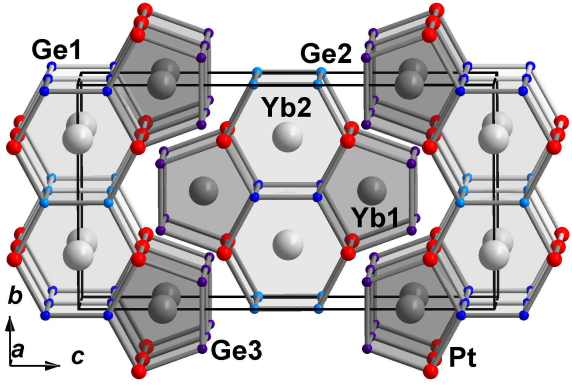}
\caption{(color online) Crystal structure of YbPtGe$_2$.}\label{fig:1}
\end{figure}

In this context the very recently studied intermetallic compound YbPtGe$_2$
deserves particular interest. YbPtGe$_2$ crystallizes with the YIrGe$_2$ type
(space group $Immm$, $a$ = 4.33715\,\AA, $b$ = 8.73518\,{\AA} and $c$ =
16.14684\,\AA.\cite{Gumeniuk-2012a} As shown in Fig.\ \ref{fig:1}, two Yb
atoms occupy the $4i$ (Yb1) and $4g$ (Yb2) sites, three Ge atoms are located
at $4j$ (Ge1), $8m$ (Ge2), and $8l$ (Ge3) sites, and one Pt atom at the $8l$
site in the unit cell. The different bonding situation suggests dissimilar
valence states between Yb1 and Yb2 site. The species on the Yb1 site, which
has a coordination polyhedron of rather small volume, may be considered as an
electronic configuration close to $4f^{13}$ (magnetic trivalent Yb). On the
other hand, the atoms on the Yb2 site within a quite large hexagonal prismatic
coordination polyhedron are expected to have the 4$f^{14}$ configuration
(non-magnetic divalent Yb).\cite{Gumeniuk-2012a}

Because of two inequivalent crystallographic sites of Yb with different
electronic configurations, YbPtGe$_2$ is considered as a multivalent charge
ordered system. Moreover, a huge drop of the magnetic susceptibility ($\chi$)
in the temperature range of 50--200 K could not be explained by the simple
valence fluctuations model.\cite{Gumeniuk-2012a} Such a drop in $\chi$ is
reminiscent of correlated semiconductors and/or semi metals such as
YbB$_{12}$,\cite{Kasaya-1983} Ce$_3$Bi$_4$Pt$_3$,\cite{Reyes-1994}
FeSi,\cite{Riseborough-2000} U$_2$Ru$_2$Sn,\cite{Rajarajan-2007} and
CeFe$_2$Al$_{10}$.\cite{Lue-PRB-CeFe2Al10-2010} However, unlike the other
Kondo semiconductors, we have not found any clear sign of a charge gap in the
electrical resistivity of YbPtGe$_2$.\cite{Gumeniuk-2012a} Therefore the
present system has been assigned as a compound with an
unconventional pseudo-gap.

Nuclear magnetic resonance (NMR) is a local probe, and therefore the NMR shift
$K \sim \chi$ provides information about the local uniform susceptibility
$\chi(q=0,\omega=0)$. The spin-lattice relaxation rate ($1/T_1 \sim \sum_q
A_q\chi^\prime(q,\omega)$) supplies information about the dynamic
susceptibility $\chi^\prime(q,w)$ and the hyperfine form factor $A_q$. In the
framework of the Korringa relation, $1/T_1$ gives a direct estimate for the
density of states (DOS) at the Fermi level if the exchange enhancement in the
metal is not too large.

Therefore NMR has been extensively and successfully applied to understand the
microscopic magnetic properties of correlated semi-metal and/or pseudo-gapped
system. While we have reported the $^{195}$Pt NMR shift data incorporated with
the bulk susceptibility data briefly,\cite{Gumeniuk-2012a} we present here
a more detailed and extended NMR study on YbPtGe$_2$.

\begin{figure}
\includegraphics[scale=0.95]{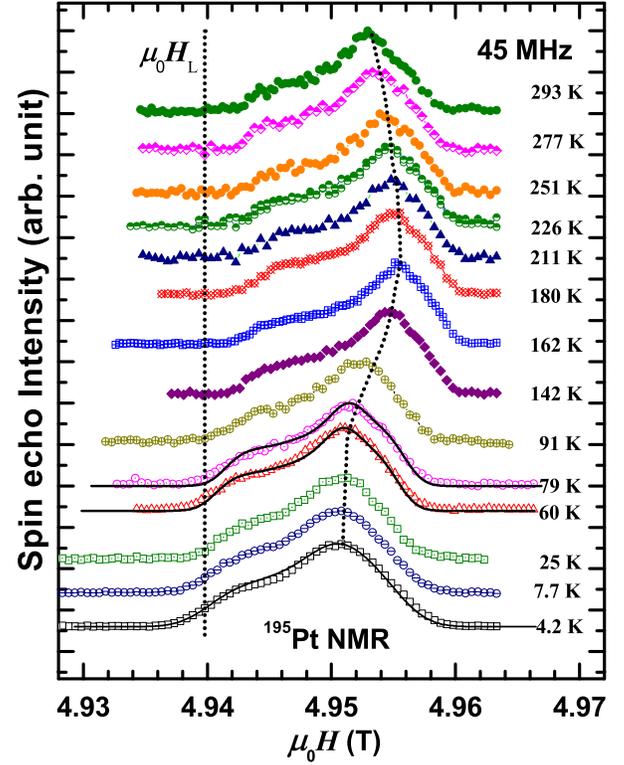}
\caption{(color online)$^{195}$Pt field sweep NMR spectra taken at 45\,MHz.
The solid lines represent theoretical simulations. The dotted curved line is
a guide to the eye for the development of the main shift component of the
shift tensor. The vertical dotted line indicates the Larmor field for
$^{195}$Pt from the non-magnetic reference YPtGe$_2$.}\label{fig:2}
\end{figure}

YbPtGe$_2$ powder samples were prepared according to Ref.\
\onlinecite{Gumeniuk-2012a}. The powder was mixed with paraffin in a small
quartz tube, subsequently heated up and shook-up, and then cured to randomize
the grains. Because of the high NMR frequency and the metallic nature of the
material such process is required to prevent the reduction of the NMR signal
due to a finite skin depth associated with radio-frequency penetration for
resonance. The first field sweep NMR measurements were carried out using a
conventional pulsed NMR technique on $^{195}$Pt nuclei (nuclear spin $I = 1/2$
and $\gamma = 9.09$\,MHz/T)\cite{Baenitz-pss-Pt-ref} in the temperature range
of 4.2\,K $\leq T \leq 295$\,K at fixed frequencies of 45 and 24\,MHz. The
$^{195}$Pt NMR shift has been determined with respect to the non-magnetic
reference compound YPtGe$_2$\cite{Francois87a} with $^{195}K \approx 0$.
Therefore the $K(T)$ values are regarded as the pure $4f$ shift component
$K_{4f}$ where a small residual temperature independent $K_0$ component is
already corrected. $K_0$ is the conduction electron contribution from the
trivalent reference. The field sweep spectra are obtained by integration over
the spin echo in the time domain at a given field. The powder spectra are
fitted with an anisotropic shift tensor to determine the components for the
$a$, $b$, and $c$ directions. The isotropic part of the Knight shift,
$^{195}K_\mathrm{iso}$ has been determined by $^{195}K_\mathrm{iso} = (K_a +
K_b + K_c)/3$. The spin-lattice relaxation rate $^{195}(1/T_1)$ has been
measured by the standard saturation recovery method in different external
magnetic fields. Additionally, we succeeded to observe the $^{171}$Yb NMR, and
the discussion of the characteristic features of the spectra is also described
below.

Figure\,\ref{fig:2} shows the $^{195}$Pt field sweep NMR spectra at different
temperatures at a fixed frequency of 45\,MHz. In all temperatures a typical
powder spectrum with three singularities at $K_\alpha$ ($\alpha$ being $a$,
$b$ and $c$) are observed. The spectra can be analyzed convincingly with the
consideration of typical random powder pattern for a spin $I = 1/2$ nuclei
with an anisotropic shift tensor caused by the orthorhombic point symmetry. We
have analyzed the spectra with three different shift values $K_a$, $K_b$ and
$K_c$. The lines in Fig.\ \ref{fig:2} indicate the best fits. With lowering
the temperature the whole spectra are shifted initially towards the high-field
side and then move towards the low-field side. However, in the temperature
range 30--1.8\,K the spectra remain unperturbed with respect to its $\alpha$
coordinate.

\begin{figure}
\includegraphics[scale=0.83]{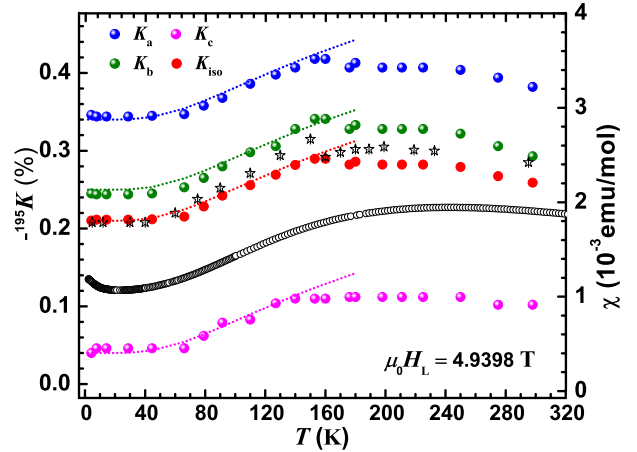}
\caption{(color online) Temperature dependence of the $^{195}$Pt NMR shift
components $^{195}K_\alpha$ ($\alpha$ = $a$, $b$, $c$) and the isotropic shift
$^{195}K_\mathrm{iso}$ in YPtGe$_2$ as estimated from the theoretical fits.
$^{195}K_\mathrm{iso}$ for 45\,MHz ($\bullet$, red) and 24\,MHz ($\star$,
black), see text. Also the bulk magnetic susceptibility\cite{Gumeniuk-2012a} is
shown ($\circ$, black). The dotted line represents the activation gap
model.}\label{fig:3}
\end{figure}

One point of interest in Fig.\,\ref{fig:2} is the absence of drastic changes
of the line profile with lowering of the temperature. This suggests that
YbPtGe$_2$ does not have any short range ordering or defects leading to a
distribution of internal fields and corresponding NMR line broadening. The
estimated $^{195}$Pt NMR shift results of YbPtGe$_2$ for the three
crystallographic axes together with the isotropic shift are shown in
Fig.\,\ref{fig:3}. The temperature dependence of the shifts in the three
directions is quite similar, except at high temperatures where $K_c$ is less
temperature dependent than $K_a$ and $K_b$. The remarkably different hyperfine
field in $c$ direction could be related to the slight lattice distortion of
this system which affect the hyperfine field. The isotropic shift
$^{195}K_\mathrm{iso}$ nicely follows the bulk
susceptibility,\cite{Gumeniuk-2012a} except for $T < 30$\,K where a small
amount of paramagnetic impurities prevails in the bulk susceptibility (``Curie
tail'') whereas the shift gets constant for $T \rightarrow 0$. We have already
shown the shift to be proportional to the susceptibility with the temperature
as an implicit parameter by a $K$--$\chi$ plot (``Clogston-Jaccarino''
plot).\cite{Clogston-pr-1964} The observed isotropic shift is related to the
magnetic bulk susceptibility by the expression,

\begin{equation}
K_\mathrm{iso}(T) = K_0 + A_\mathrm{hf} \chi(T)/N_A\mu_B,
\end{equation}

where $A_\mathrm{hf}$ is the hyperfine coupling constant, $\chi$ is the bulk
magnetic susceptibility and $K_0$ is a residual temperature independent shift
contribution.  We obtain the hyperfine coupling constant of $A_\mathrm{hf} =
-7.261$\,kOe/$\mu_B$ and $K_0 = -0.21$\,{\%}. In the $K$--$\chi$ plot at high
temperatures a change of slope is observed, indicating the change of the
hyperfine coupling constant. Finally it should be mentioned that the shift is
negative as expected for Yb-based compounds by employing a simple conduction
electron polarization model.\cite{carter1977metallic} The finite $K_0$ value with respect to the non magnetic homologue YPtGe$_2$ is due to a conduction electron contribution (positive) and/or a chemical shift contribution (negative).

Following the bulk magnetic susceptibility, the negative shift increases
rather sharply with increasing temperature in the range from 35\,K to 160\,K.
Subsequently, the shift exhibits a broad maximum and thereafter decreases
slowly following a Curie-Weiss law. The present results are reminiscent of
correlated semi-metal systems such as YbB$_{12}$,\cite{Kasaya-1985}
U$_2$Ru$_2$Sn,\cite{Rajarajan-2007} Ce$_3$Bi$_4$Pt$_3$ \cite{Reyes-1994},
CeFe$_2$Al$_{10}$,\cite{Lue-PRB-CeFe2Al10-2010} and
U$_3$Bi$_4$Ni$_3$.\cite{Baek-2009} The sharp decrease of the shift with a
residual shift value of around $-0.21$\,{\%} may be interpreted as the opening
of a pseudo-gap at the Fermi energy. To get an idea of the gap size of this
system, we have fitted the $^{195}K^\alpha$ data with a simple activation gap
model $K^\alpha_0 + A^\alpha \times \exp(-\Delta^\alpha/k_BT)$. The fits
result in similar gap values for all three direction with gap sizes of
$\Delta^\alpha/k_B \approx 200$\,K. It should be mentioned that the residual
shift values $K^\alpha_0$ are rather different, which indicates the anisotropy
of the electronic band structure of the orthorhombic system.

\begin{figure}
\includegraphics[scale=0.81]{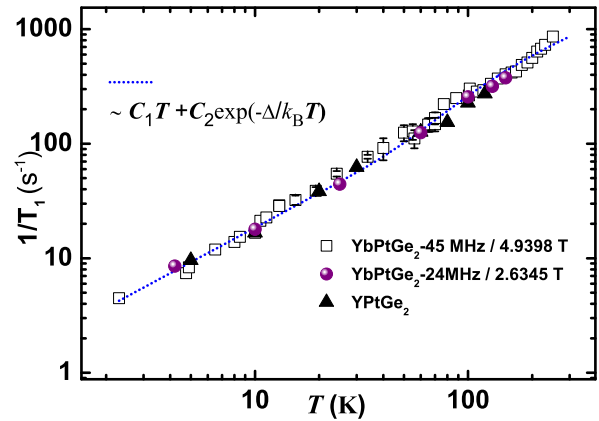}
\caption{(color online) $^{195}$Pt spin-lattice relaxation rate \textit{vs}
$T$ plot at frequencies 24 and 45\,MHz for YbPtGe$_2$. In addition,
$^{195}$(1/$T_1$) of the non-magnetic reference compound YPtGe$_2$ is
plotted. The dashed line represents a pseudo-gap fit (see
text).}\label{fig:4}
\end{figure}
\begin{figure}
\includegraphics[scale=.85]{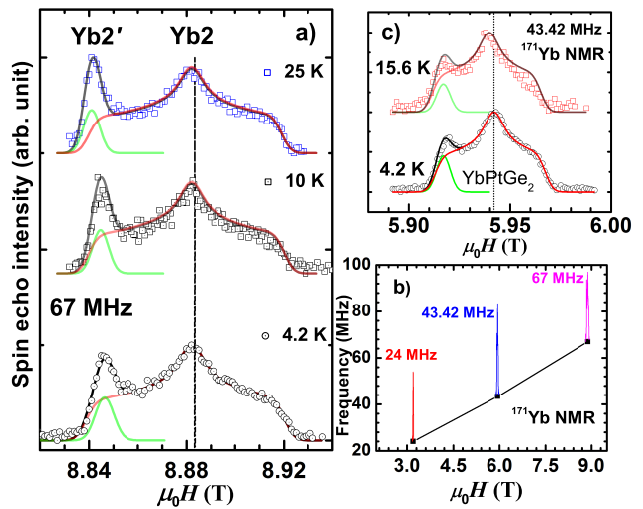}
\caption{(color online) $^{171}$Yb field sweep NMR spectra at different
frequencies for YbPtGe$_2$. a) and c) shows the $^{171}$Yb spectra at 67\,MHz
and 43.42\,MHz together with the calculation by a model with two Yb2 sites
with different anisotropy, respectively. b) $^{171}$Yb field sweep NMR
spectra and frequencies as a function of resonance field at 4.2\,K. The solid
line provides an average $\gamma$ for Yb.}\label{fig:5}
\end{figure}

As already mentioned, 1/$T_1$ is very sensitive to the DOS at the Fermi
energy. For a simple metal, the following relation is obtained,
\cite{TOU-Ishida-Kitaoka,Baenitz-NMR-FM,Rajib-Korringa-law}

\begin{equation}
1/T_1\propto T\sum_q A_q\chi^\prime(q,\omega) \propto \mathcal{K}^2(\alpha) K^2T
\label{equation2}
\end{equation}

where $\mathcal{K}(\alpha)$ is an enhancement factor which depends on $\alpha
\propto \chi^\prime(q \neq 0) / \chi\prime(q = 0)$ according to the Stoner
theory. \cite{TOU-Ishida-Kitaoka,Baenitz-NMR-FM,Rajib-Korringa-law} If the
enhancement factor $\mathcal{K}(\alpha) = 1$, then Eq.\,\ref{equation2}
becomes

\begin{equation}
1/T_1TK^2 = S_0
\label{equation3}
\end{equation}

which is known as the Korringa relation and $S_0$ is the Korringa product. To
see the effect of this pseudo-gap on the dynamical susceptibility
$\chi^\prime(q,\omega)$, we have also investigated the spin-lattice relaxation
rate $^{195}(1/T_1$), which is a direct measure of the DOS. With the opening
of a gap, $^{195}(1/T_1)$ should decrease drastically and, in a simple
activation gap model, it should decrease exponentially, i.e.\ ($^{195}(1/T_1)
\propto \exp(-\Delta/k_BT$). For spin $I = 1/2$ nuclei the magnetization
recovery curve should be followed by the equation

\begin{equation}
1-M(t)/M(0) = A \exp (-t/T_1)
\label{Eq4}
\end{equation}

where $M(t)$ is the magnetization at a time $t$ after the saturation pulse and
$M(0)$ is the equilibrium magnetization.

In our experiments all magnetization recovery curves follow Eq.\ \ref{Eq4} and
by fitting $^{195}(1/T_1$) was estimated. Figure\,\ref{fig:4} shows the
$^{195}(1/T_1$) \textit{vs} $T$ plot at two different fields for YbPtGe$_2$
and the reference compound YPtGe$_2$. One can immediately recognize that the
temperature dependence of $^{195}(1/T_1$) in YbPtGe$_2$ is nearly identical to
that of non-magnetic YPtGe$_2$. The dashed line is a fit with a pseudo-gap
model\cite{Rajarajan-2007} where only a small fraction of states are gapped
and the rest stays non-gapped. Such a scenario implies a large number of
non-gapped states which is less realistic. Another reason might be that the
magnetic fluctuations are filtered out by the hyperfine form factor $A_q$ at
the Pt site.

In most Kondo insulators, the opening of the gap is evidenced by a
considerable drop in the $1/T_1$ \textit{vs} $T$ plot, such as in
U$_2$Ru$_2$Sn and FeSi. This further indicates that the Yb $4f$-electron
dynamical spin and/or charge fluctuations do not at all influence the
$^{195}$Pt nuclei, and, seemingly $^{195}(1/T_1$) is insensitive to the spin
and/or charge excitations. Most probably the uniform susceptibility deviates
strongly from the complex $q$-dependent dynamic susceptibility
$\chi^\prime(q,\omega)$. It is worth to mention here that there exists no
clear evidence of the gap in electrical resistivity data, which behave like a
simple metal as in the $^{195}(1/T_1$) results. Therefore, from the
$^{195}(1/T_1$) data we rule out a Kondo-insulator scenario for YbPtGe$_2$.

The aforementioned model with two different Yb species (nearly divalent Yb on
Yb2 site, nearly trivalent magnetic Yb on Yb1 site) opens up the opportunity
to probe non-magnetic Yb ion on the Yb2 site by NMR, which is a rare case
($^{171}$Yb, $I = 1/2$, $\gamma = 7.4987$\,MHz/T). We focused our NMR study on
selected frequencies (fields) as indicated in Fig.\,\ref{fig:5}. The
$^{171}$Yb NMR line was obtained at several frequencies. In Fig.\,\ref{fig:5}b
the evolution of the resonance frequency of the center line (illustrated with
a spectra) with the resonance field at 4.2\,K is shown. The slope of the line
roughly corresponds to the textbook $\gamma$ value for $^{171}$Yb (average
value 7.4575 MHz/T). The shape of the field sweep NMR spectra clearly reveals
the orthorhombic site symmetry (Fig.\ \ref{fig:5}a and c). The line position
itself is very close to the Larmor field of $^{171}$Yb which further
corroborates that we are probing the non-magnetic ions on the Yb2 site of the
YbPtGe$_2$ lattice (cf.\ Fig.\,\ref{fig:5}b). In contrast to Pt, the Yb2 site
seems to be weakly coupled to the magnetic Yb1 site and therefore no sizable
temperature dependence of the shift is found. The line is consistently
(67\,MHz and 43.42\,MHz) simulated by taking three temperature-independent
weak shift tensor components $K_\alpha = +0.5$\,{\%} (left), 0.01\,{\%}
(middle) and $-0.48$\,{\%} (right).

Surprisingly, we observe an additional $^{171}$Yb NMR line superimposed to the
main Yb2 line. This Yb2$^\prime$ line has also only a weak and
temperature-independent shift $K^\prime\cong$ 0.5\,{\%} which directly implies
that its origin is non-magnetic Yb$^{2+}$. Most likely this additional
Yb2$^\prime$ signal is not a foreign phase because it is centered on the left
singularity of the main spectrum. We believe that the origin of this line is a
partial texture in the powder sample which amplifies the signal in this
particular orientation. Such an effect is common in NMR on powder
samples.\cite{carter1977metallic}

To sum up, our present Pt-NMR shift investigations tracks down microscopically
the bulk susceptibility, evidencing the exotic nature of YbPtGe$_2$. We found
an isotropic gap from the shift analysis with $\Delta/k_B \approx 200$\,K. The
$^{195}(1/T_1$) results are in good agreement with resistivity data, which
shows that there is no charge gap, clearly ruling out a Kondo-insulator
scenario.

Further, the dynamical spin fluctuations of the Yb $4f$-electrons seem to be
filtered out at the Pt sites, so that $^{195}(1/T_1$) behaves like a simple
metal similar to YPtGe$_2$. It could be speculated that this is due to a
complex $q$-dependence of $\chi(q,\omega)$. However, neither the relation
$^{195}(1/T_1TK^2)$ = constant (simple metallic) nor $^{195}(1/T_1TK)$ =
constant (weakly magnetic) are valid in this system. In good agreement with
the suggested scenario\cite{Gumeniuk-2012a} of two different Yb species
(nearly divalent Yb on Yb2 site, nearly trivalent magnetic Yb on Yb1 site),
the NMR gives good evidence for non-magnetic Yb species on the Yb2 site.

So far, in our NMR investigations, the magnetic Yb$^{3+}$ ion on the Yb1 site
could not be found. Magnetic Yb is observed only in the rare cases of some
valence-fluctuating systems like YbAl$_2$ and
YbAl$_3$,\cite{171Yb-NMR-YbAl2-YbAl3} or YbB$_{12}.$\cite{Ikushima2000274}
Here, the residual NMR shift due to the intermediate valence of Yb is very
large and not easy to detect.\cite{171Yb-NMR-YbAl2-YbAl3,Ikushima2000274} The
on-site interaction in these Yb systems is well understood and allows the
calculation of the shift from the Yb-dominated bulk
susceptibility.\cite{Calculation-NMR-shift} Our NMR data clearly reveal the
presence of well-ordered Yb$^{2+}$ species with orthorhombic site symmetry.
This finding is a microscopic proof for the proposed multivalent
charge-ordered system scenario.

We acknowledge Prof.\ Hiroshi Yasuoka for stimulating discussions. R.\ Sarkar
is grateful to the DFG for the financial support with grant no.\ SA\,2426/1-1.

\bibliographystyle{apsrev}
\bibliography{Sarkar}

\begin{thebibliography}{30}
\bibitem[{Cal()}]{Calculation-NMR-shift}
\bibinfo{note}{Assuming a hyperfine form factor $A^\mathrm{hf}_\mathrm{Yb} =
  115\,\mathrm{T}/\mu_B$ leads to a shift of $\approx 67$ percent for the
  $^{171}$Yb$^{3+}$ line in YbPtGe$_2$: $H_\mathrm{res}^\mathrm{Yb} = H_L / (1
  + \chi A^\mathrm{Yb}_\mathrm{hf}) \approx H_L / (1 + 0.495) \approx 0.67
  H_L$, where $H_L$ is the Larmor field of $^{171}$Yb and $\chi$ the
  susceptibility in units of $[\mu_B/\mathrm{T}]$.}
\end{thebibliography}
\end{document}